\documentclass{article}
\usepackage{spconf,amsmath,graphicx}
\usepackage{orcidlink}
\usepackage{epsfig}
\usepackage{multirow}
\usepackage{hyperref}
\usepackage{cleveref}
\usepackage{caption}
\usepackage{subcaption}
\usepackage[para,online,flushleft]{threeparttable}
\usepackage{colortbl}
\usepackage{times}
\usepackage{booktabs}

\title{Strong Baseline and Bag of Tricks for COVID-19 Detection of CT Scans}
\name{Chih-Chung Hsu*, Chih-Yu Jian, Chia-Ming Lee, Chi-Han Tsai, and Sheng-Chieh Dai}
\address{Institute of Data Science, Department of Statistics, \\
*Center of Data Science, Miin Wu School of Computing, \\
National Cheng Kung University,\\
No.1, University Rd. Tainan City, Taiwan \\
cchsu@gs.ncku.edu.tw}
\begin{document}
%
\maketitle
\begin{abstract}

This paper investigates the application of deep learning models for lung Computed Tomography (CT) image analysis. Traditional deep learning frameworks encounter compatibility issues due to variations in slice numbers and resolutions in CT images, which stem from the use of different machines. Commonly, individual slices are predicted and subsequently merged to obtain the final result; however, this approach lacks slice-wise feature learning and consequently results in decreased performance. We propose a novel slice selection method for each CT dataset to address this limitation, effectively filtering out uncertain slices and enhancing the model's performance. Furthermore, we introduce a spatial-slice feature learning (SSFL) technique\cite{hsu2022} that employs a conventional and efficient backbone model for slice feature training, followed by extracting one-dimensional data from the trained model for COVID and non-COVID classification using a dedicated classification model. Leveraging these experimental steps, we integrate one-dimensional features with multiple slices for channel merging and employ a 2D convolutional neural network (CNN) model for classification. In addition to the aforementioned methods, we explore various high-performance classification models, ultimately achieving promising results.
\end{abstract}
\section{Introduction}
\label{sec:intro}

To diagnose SARS-CoV-2 (COVID-19) infection, physicians rely on examining patients' lung CT images. However, a single patient's CT scan may comprise hundreds of images, rendering manual analysis inefficient, particularly when physicians must assess dozens or even hundreds of patients daily. The advent of deep learning models in recent years has facilitated their application as aids in clinical diagnosis and examination. Compared to X-rays, CT scans offer more detailed and comprehensive image information, enabling the clear visualization of solid organ interiors and assisting physicians in determining infection severity. Moreover, high-quality 3D images generated from CT scans benefit deep learning models that demand intricate feature details and image information.

Many studies have employed deep learning for CT image analysis in recent years. Guhan et al. \cite{guhan2022automated} utilized the k-means algorithm to extract regions of interest in CT images and harnessed the gray level co-occurrence matrix (GLCM) to extract texture features. They subsequently applied both machine learning and deep learning models to classify the presence or absence of COVID-19 symptoms. Experimental results revealed superior performance from the deep learning model (accuracy 0.99) compared to the traditional machine learning model (accuracy 0.97). Zhang et al. \cite{SaiZHang-2022}, building upon the VGG architecture and employing the GlobalMaxPool 2D Layer, achieved enhanced performance in AUC, sensitivity, specificity, and accuracy evaluation metrics relative to traditional CNN models and vision transformer (ViT) models \cite{dosovitskiy2020image}, with the highest test result of AUC 0.9744 and accuracy 0.9412. These studies demonstrate the efficacy of deep learning models in COVID-19 CT image classification tasks.

Kolliaz et al. proposed the COVID-19-CT-DB dataset \cite{kollias2022ai, arsenos2022large,kollias2021mia,kollias2020deep,kollias2020transparent,kollias2018deep}, which encompasses a vast amount of labeled COVID-19 and non-COVID-19 data to address the issue of deep learning models requiring extensive data for training. However, CT images exhibit varying resolutions and slice numbers depending on the machine used, presenting a challenge to the architecture of deep learning models. Chen et al.\cite{chen2021adaptive} introduced two methods to determine the importance of CT image slices. In the 2D method, they proposed the Adaptive Distribution Learning with Statistical hypothesis Testing (ADLeaST) method, which integrates statistical analysis with deep learning for COVID-19 CT image classification. This approach maps CT images to a specific distribution and excludes slices devoid of clear lung tissue, thereby allowing for the assessment of each slice's importance and yielding stable and interpretable prediction results. Nonetheless, 2D image detection remains vulnerable to the influence of positive slices without overt symptoms, as these slices are randomly sampled from the dataset during training. The 3D method incorporates self-attention structures (Within-Slice-Transformer, Between-Slice-Transformer) into the 3D CNN architecture. However, this approach is hindered by the 3D framework, leading to insufficient training samples and overfitting issues due to the complex model architecture.

Furthermore, in \cite{hsu2022}, the authors proposed an efficient spatial-slice feature learning (SSFL) method to address the aforementioned CT image challenges during the 2nd COVID-19 CT competition. We employed a conventional and high-performance 2D CNN model for feature learning across all slices. Subsequently, the Swin-transformer processed the semantic feature embedding results generated by the 2D CNN to facilitate feature learning between slices. Moreover, we introduced an essential slices set algorithm to tackle difficult-to-diagnose slices and to select CT slices containing significant image information, thereby enhancing the performance of SSFL. Comprehensive experiments demonstrated that our proposed SSFL method and slice selection approach yielded excellent performance scores and stable classification and detection results.

\section{Methodology}
\label{sec:method}

We employed the approach proposed in \cite{hsu2022} as the baseline and followed by improving the feature aggregation and enhancement for better performance. This methodology primarily encompasses the previously mentioned SSFL and slice selection techniques, as well as a newly introduced multi-model ensemble method. In the subsequent sections, we will describe the technical and implementation details. First, we will reveal the details of the SSFL for better preprocessing purposes, followed by introducing more powerful and essential methods to extract representative features from CT scans to meet high accuracy of COVID-19 detection.

\begin{figure}[htb]
\begin{minipage}[b]{1.0\linewidth}
  \centering
  \centerline{\includegraphics[width=8.5cm]{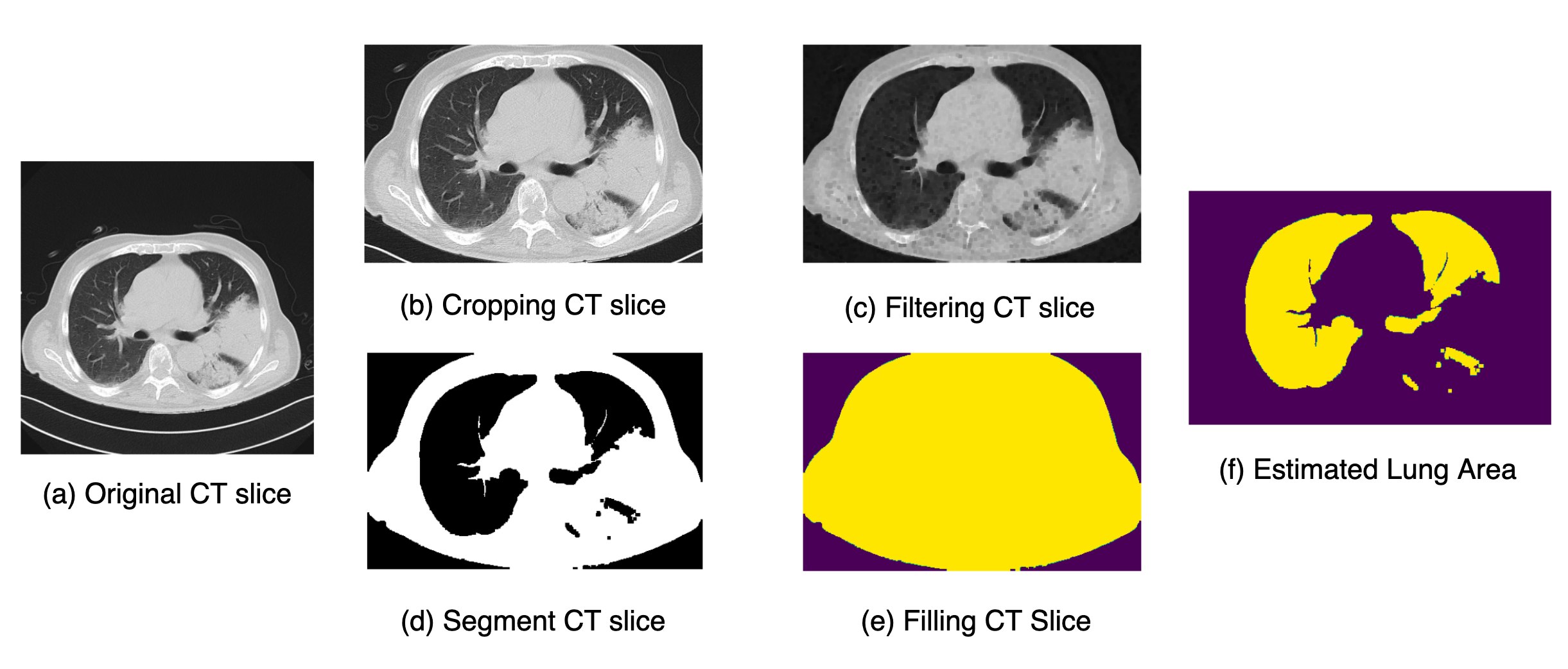}}
\end{minipage}
\caption{CT Image Preprocessing Flowchart. (a) to (f) represent the complete single-slice processing procedure, including the original slice, cropped slice, filtered slice, slice segment map, dilated filtered slice, and the resulting image.}
\label{fig:lung}
\end{figure}

\subsection{An algorithm for selecting important slices}
\label{sec:lung_processing}
During the slice region processing stage, given a label $y$ and $n$ slices, we define $CT={\mathbf{Z}i}^{n}{i=1}$, to obtain the slice image $CT^*={\mathbf{Z}i}^{e}{i=s}$, which encompasses the largest area of lung tissue within a range of slices, where the start index is $s$ and the end index is $e$. The initial step involves employing a cropping method to eliminate the background of the CT image slices, resulting in a cropped region $CT_{crop}={(\mathbf{Z}{crop}^{(i)})}^{e}{i=s}$. This technique filters out non-lung tissue areas and reduces computational complexity. Subsequently, the minimum filtering operator with a window size of $k\times k$ is applied to mitigate further noise introduced by the CT scanner. These steps can be defined as:

\begin{equation}
    \label{eq:filter01}
    \mathbf{Z}_{filter} = Filter(\mathbf{Z}_{crop}, k).
\end{equation}

The above formula can determine the segmentation map $\mathbf{Map}$ of the filtered slices by a threshold $t$:

\begin{equation}
    \label{eq:seg}
    \mathbf{Map}[i,j] = 
    \begin{cases}
    0,\,\text{if}\,\mathbf{Z}_{filter}[i,j] < t\\
    1,\,\text{if}\,\mathbf{Z}_{filter}[i,j] >= t
    \end{cases}
\end{equation}

 To find the lung tissue region in the CT scan, we used the binary dilation algorithm \cite{enwiki:1082436538} to perform a filling check and obtain the filling result $\mathbf{Map}_{filling}$z Here, $i$ represents the width index of the slices, and $j$ represents the height index of the slices. The detailed illustration can be found in \cref{fig:lung}. The difference between the segmentation and filling maps represents the lung tissue region. The above method can be summarized as the following formula:

 \begin{equation}
    \label{eq:area}
    Area(\mathbf{Z}) = \sum_i\sum_j\mathbf{Map}_{filling}(i,j) - \mathbf{Map}(i,j).
\end{equation}

After the above screening, we can finally obtain a range where $s$ and $e$ denote the starting and ending indexes, respectively, and $n_c$ is the constraint of the number of slices for a single CT scan.

\begin{equation}
    \label{eq:area}
    \begin{split}
    \mathop{\arg\max}_{s,\,e}{\sum^e_{i=s}Area(\mathbf{Z}_i)}, \\
    \text{subject to } e-s \leq n_c,
    \end{split}
\end{equation}

Finally, during training, the slice range $CT^*$ will be fed into the model as a sub-training set. The binary cross-entropy loss function will be used as the loss function for binary classification in this experiment. Let $NC$ be the total number of selected slices, and $p_i=F_{2D}(\mathbf{X}i;\theta{2D})$, where $\theta_{2D}$ is the trainable parameters of the 2D CNN model.

\begin{equation}
    \label{eq:bce}
    Loss_{bce} = -\frac{1}{NC}\sum^N_{i=1}y_i\cdot\log(p_i)+(1-y_i)\log(1-p_i),
\end{equation}

\subsection{ECA-NFnet and Spatial-Slice Feature Learning}
\label{ssfl}
This section employs two types of two-step networks, i.e., ECA-NFnet \cite{brock2021high} and EfficientNet\cite{effnet} networks, for model construction. The following paragraphs will provide an overview of the architecture of these two types of models and the advantages of each network.

The model characteristic of \textbf{ECA-NFnet} is that it does not include any normalization layers but only uses Weight Standardized convolutions. This modification can accelerate the training process, and the SE-Resnet module inside the NFNet\cite{brock2021high} is replaced with the Efficient Channel Attention (ECA) module. This allows the model to focus more on important channel features, while the ECA module is more computationally efficient than some Attention modules. Thus, it improves the performance and efficiency of the model.

Next, EfficientNet\cite{effnet} is used for slice-level feature learning, and only partial slices of CT are used for learning. Before the start of the second stage, we send all slices to EfficientNet to extract embedding features so that only important slice features are retained in the CT image. Then, we send the embedding feature data to the next stage network for classification. The next stage network is a hybrid of 1dcnn and 2DCNN, referred to as \textbf{eff-mix-conv}. First, we use 1dcnn to filter the embedding feature, and then stack the filtered embedding feature to change the data dimension to 3D. Finally, we use 2DCNN for the classification task.

\section{Experiments}
\subsection{Experimental Settings}
In this competition, we used a total of 8040 COVID-19-CT-DB data\cite{kollias2022ai, arsenos2022large,kollias2021mia,kollias2020deep,kollias2020transparent,kollias2018deep}, and the detailed Train, Valid, and Test data information can be found in \cref{tab:dataset_subgroup}. After careful inspection, we found that some of the training data contained partially or completely non-horizontal CT scans of the lungs. We manually removed these data to reduce training variability. Except for the aforementioned exceptions, all other preprocessing methods were performed using the methods proposed in \cref{sec:lung_processing}.
\begin{table}[!ht]
    \centering
    \caption{Performance comparison between the proposed method and the baseline.}
    \begin{tabular}{l c c c c} 
      \toprule 
       Data type& COVID-19 & Non-COVID-19&   total \\
      \midrule
      Train     &       922&         2106&     3028\\
      \midrule
      Valid     &       235&          469&      704\\
      \midrule
      Test      &         -&            -&     4308\\
      \midrule
      \bottomrule
    \end{tabular}
    \label{tab:dataset_subgroup}
\end{table}
We mainly used Sensitivity, Specificity, and F1-score in the experimental process for model evaluation, with the first two indicators often used in statistics or clinical trials. Sensitivity can represent the model's sensitivity to COVID-19 positive samples, and a higher value of this indicator typically indicates a more reliable detection capability of the model for positive diseases. On the other hand, specificity can be used to determine whether the model will misclassify negatives as positives, and a higher value of this indicator means that the false-positive results will be lower. F1-score is a metric used to determine the accuracy of a binary classification model. It combines the harmonic mean of Precision and Recall (Sensitivity). Precision evaluates the model's ability to correctly predict true positives, while Recall is conceptually the same as Sensitivity.

\textbf{ECA-NFnet.} In this experiment, Coarse Dropout\cite{devries2017improved} was applied with a $p$ value of 0.2 to preserve better spatial structural features in the feature maps. The Adam\cite{Adam} optimizer was used with a learning rate of 1e-4 and a weight decay of 5e-4, and the threshold was set to 0.5. All training images were resized to [$3\times384\times384$]. Cross-validation was also used to ensure model stability and check performance during the experiment. The model is referred to as ECA-NFnet-CV in the subsequent discussions.

\textbf{Eff-mix-conv.} This two-step model consists of two architectures. The first architecture uses EfficientNet-b3a\cite{rw2019timm} as the backbone and adds a fully connected layer with a channel output of 224 to reduce computational complexity. During the training phase, 16 slices are randomly selected as a small batch. The second architecture has two different designs. The first design uses a 1D convolution layer with a kernel size of 1, an input size of 224 (embedding size), and an output size of 100. The feature map output from this layer is stacked three times, with a shape of [$3\times100\times100$], and used as the input size for the ResNet18 model. This design is referred to as \textbf{eff-mix-conv-E}. The second design sets the input size of the 1D convolution layer to 100, where 100 represents the number of slices and is referred to as \textbf{eff-mix-conv-S}. All \textbf{eff-mix-conv} models use the Adam optimizer with a learning rate of 1e-4 and a weight decay of 5e-4. The threshold is set to 0.5. All training images are resized to [$3\times256\times256$].

\subsection{Performance evaluation}
From the evaluation metrics in \cref{tab:result1}, it can be observed that the performance of this year's model is better than the architecture proposed by our team last year. Among them, the eff-mix-conv model architecture obtains a higher F1-score, while the ECA-NFnet achieves the highest sensitivity. As our model adopts the method of randomly selecting slices for slice-level feature recognition training in the first stage, there may be slight fluctuations in the evaluation metrics. In order to obtain stable evaluation metrics, we use an internal ensemble method to run each model framework 10 times and take the average, to obtain stable and reliable metrics.

\begin{table}[!ht]
    \centering
    \caption{}
    \begin{tabular}{l c c c c} 
      \toprule 
                                                 &    SE &    SP & Macro-F1 \\
      \midrule
      Previous year's method \cite{hsu2022}      & 0.794 & 0.921 &     0.861\\
      \midrule
      ECA-NFnet-CV (Mean 10)                     & 0.885 & 0.961 &     0.927\\
      \midrule
      eff-mix-conv-E                             & 0.865 & 0.963 &     0.922\\
      eff-mix-conv-S                             & 0.842 & 0.965 &     0.913\\
      \midrule
      \bottomrule
    \end{tabular}
    \label{tab:result1}
\end{table}

\section{Conclusions}
This paper improves upon the method proposed last year\cite{hsu2022} by utilizing special CT slices image preprocessing to enhance the texture intensity of each slice, as well as using a two-stage model to solve the problem of inconsistent numbers of CT image slices per group. From the experimental results, it can be observed that the network architecture used by our team this time can significantly improve the performance compared to the method used last year.

\bibliographystyle{IEEEbib}
\bibliography{refs}
\end{document}